\documentstyle[12pt,aps,prl]{revtex}
\begin{document}

\draft

\title{Aging Effects in a Lennard-Jones Glass}

\author{Walter Kob }
\address{ Institut f\"ur Physik, Johannes Gutenberg-Universit\"at,
Staudinger Weg 7, D-55099 Mainz, Germany}

\author{Jean-Louis Barrat }
\address{ D\'epartement de Physique des Mat\'eriaux \\
Universit\'e Claude Bernard and CNRS, 69622 Villeurbanne Cedex, France}

\date{\today}

\maketitle

\begin{abstract}
Using molecular dynamics simulations we study the out of equilibrium
dynamic correlations in a model glass-forming liquid. The system is
quenched from a high temperature to a temperature below its glass
transition temperature and the decay of the two-time intermediate
scattering function $C(t_w,t+t_w)$ is monitored for several values of
the waiting time $t_w$  after the quench.  We find that $C(t_w,t+t_w)$
shows a strong dependence on the waiting time, i.e. aging, 
depends on the temperature before the quench and, similar to the case
of spin glasses, can be scaled onto a master curve.

\end{abstract}

\pacs{PACS numbers: 61.43.Fs, 61.20.Lc , 02.70.Ns, 64.70.Pf}


Thanks to a remarkable combination of  experimental, theoretical and
numerical work, in the last few years considerable progress has been
made in the understanding of the dynamical properties of disordered
systems such as spin glasses \cite{Vincent96}. In particular, the
importance of the so called aging phenomenon, i.e. the out of
equilibrium evolution of a system quenched into a glassy state, has
been realized. This phenomenon, which is well known experimentally in
structural (e.g. polymer) glasses, was shown to display  some universal
scaling features, common to spin and structural glasses. Subsequently,
several theoretical interpretations of the phenomenon  were put
forward, involving either phenomenological ``trap''~\cite{traps}   or
``coarsening''~\cite{coarsening} models,  or the solution of dynamical
equations for disordered systems in the mean-field limit
\cite{Bouchaud97}.  Interestingly, the dynamical equations describing
these models turn out to have a structure which is extremely similar to
the structure of the so called mode-coupling equations, that were
developed by G\"otze, Sj\"ogren and coworkers in order to describe the
glass transition singularity in {\it structural} glass-formers
\cite{Gotze92}. Again, this similarity at the formal level points
towards a possible connection between the slow dynamics in structural
and spin glasses, a possibility that has recently attracted much
interest~\cite{Kirkpatrick89,Parisi97}. In this work, we investigate the
nonequilibrium dynamics of a simple structural glass-former, seeking
for evidence of an ``aging'' behavior similar to what is observed in
spin glasses.  Although the method we use (Molecular Dynamics (MD)
simulations) is limited to relatively short time scales, typically
$10^{-8}$s, and high quenching rates, it has proven to be very useful
in assessing the applicability of mode coupling theory to structural
glass-formers \cite{glass_transition_MD}.  Hence we believe that the
universal features of nonequilibrium slow dynamics, if they exist,
should already appear at such short time scales.

The model glass-former we study in this work is a binary mixture of
particles interacting through Lennard-Jones potentials. This system has
been studied extensively by one of us \cite{Kob}, and we refer to these
papers for a detailed description of the model and of its
equilibrium properties.  For the present purpose, it will be sufficient
to say that the {\it equilibrium} dynamics of the model has been shown
to be well described -on the time scale of MD simulations- by ``ideal''
mode coupling theories (MCT)~\cite{Gotze92}, with a critical
temperature (in reduced Lennard-Jones units) $T_c\simeq 0.435$~\cite{Kob}. 

Our aim here is to study the nonequilibrium properties of this well
characterized model. Our strategy is the following: Starting from
an {\it equilibrium} system  at some initial temperature $T_i> T_c$, 
we instantaneously ``quench'' the system by rescaling particles
velocities to a final temperature $T_f$. The system is subsequently
allowed to evolve {\it at constant temperature} for $5\cdot 10^6$ time
steps, which corresponds to $10^5$ reduced Lennard-Jones time units
(typically  $10^{-8}$s for a real system). This procedure is then
repeated  for several independent starting configurations (typically 10)
in order to improve the statistics of the results.

The evolution of the system towards equilibrium can in principle be
monitored by using ``one-time'' observables, such as the energy or the
pressure.  Unfortunately these observables are rather insensitive to
the very slow evolution of the system that takes place after the
quench. After a fast ``equilibration'' period of several thousand LJ
time units, they essentially level off, as can be see in the time
dependence of the potential energy $e_{pot}$ (see inset of
Fig.~\ref{energy_vs_t}). A naive conclusion would be that the system is
at equilibrium. This is, however, not correct, since a careful
examination of the data shows that, for $T_i=5.0$ and 0.8,
$e_{pot}$ shows a power-law dependence on time
(Fig.~\ref{energy_vs_t}), with an exponent around 0.144, and that such
a functional form is also compatible with the data of $T_i=0.466$ at
long times. Since the exponent is small, the data is also compatible
with a logarithmic dependence on time, but since it was recently shown
that for a soft sphere system $e_{pot}(t)$ shows a power-law
dependence~\cite{Parisi97}, this functional form seems to be more
appealing. Thus we find that this ``one-time'' quantity shows indeed
only a small sensitivity on the nonequilibrium evolution of the system,
which is in agreement with the theoretical prediction~\cite{Bouchaud97}
and was also found, e.g., in Ref.~\cite{andrejew96}.

The nonequilibrium dynamics is much better characterized by the study
of two-time correlations functions, of the form $C_A(t_w,t+t_w)=
\langle A^{\star}(t_w) A(t_w+t)\rangle$, where $A$ is a microscopic
observable, and $t_w$ is the ``waiting time'', i.e. the time elapsed
after the quench, which takes place at $t=0$. The brackets refer here
to an equilibrium average over the initial configurations, at
temperature $T_i$.  In this work, we focus on the case where $A=\exp(i
{\bf q}\cdot {\bf r}_i(t))$, where ${\bf r}_i$ is the position of atom
$i$ and ${\bf q}$ is a wave-vector.  Hence the correlation function we
compute is simply the nonequilibrium generalization of the usual
incoherent scattering function:

\begin{equation}
C_q(t_w,t_w+t) = {1\over N} \left< \sum_{i=1}^N 
\exp\left(i {\bf q}\cdot
\left({\bf r}_i(t_w+t)-{\bf r}_i(t_w) \right) \right) \right> \qquad.
\label{cdef}
\end{equation}
Figure \ref{C1} displays such correlations functions, for $T_i=5.0$ and
$T_f=0.4$. The evolution of the curves as $t_w$ increases clearly shows
that the system does not reach equilibrium within the time window
explored in the simulation. A striking fact is that if one attempts to
extract a ``relaxation time'' $t_r$ from $C_q(t_w,t_w+t)$, this relaxation
time will be a rapidly increasing function of $t_w$. The larger the
waiting time, the longer it takes the system to forget the initial
configuration. This behavior is quite typical of ``aging''
phenomena~\cite{Bouchaud97}.  Although the time scale explored here is
quite atypical in the study of such phenomena, the basic observation
is  similar to what can be seen experimentally on much longer time
scales.

Correlation functions such as those displayed in Fig.~\ref{C1} depend
on a number of parameters that can be varied independently. $q$, $t_w$
and $t$ are explicit arguments in Eq.~(\ref{cdef}), but the initial and
final temperature are also implicitly present. $T_i$ determines the
ensemble average, and  $T_f$ the dynamics after the quench.  In the
following, we concentrate on the results obtained for a value of $q$
that corresponds to the main peak in the structure factor of the fluid,
$q=7.2$ (in Lennard-Jones units)~\cite{Kob}. We also focus on quenches
to a final temperature $T_f=0.4$. At this temperature, the relaxation
time of the system, estimated by extrapolating equilibrium data from
higher temperatures using a Vogel-Fulcher law, will be of order
$\tau_e(T=0.4) \sim 10^7$ time units, much larger than our simulation
times~\cite{Kob}. We have also performed similar calculations for other
wave-vectors and other values of $T_f$, but the results do not differ in
any essential way from those presented here~\cite{barrat97}.

In order to rationalize the results obtained for various values of the
remaining parameters ($T_i$,$t$ and $t_w$), we use the popular and
intuitive picture \cite{energy_landscape} that describes the whole
system as a point evolving within a complex multidimensional (free)
energy landscape. The system starts with a high average kinetic energy
corresponding to the temperature $T_i$.  It is subsequently quenched
instantaneously, so that it will be ``trapped'' (by entropic or
energetic barriers) in a configuration typical of temperature $T_i$.
The following nonequilibrium evolution will  bring the system closer to
configurations characteristic of $T_f$, which might or might not be
reached on the time scale of the simulation.

Based on this type of picture, we can attempt to get some insight into
the nonequilibrium dynamics for systems quenched to the same $T_f$,
starting from different values of $T_i$. For clarity, let us consider
the case were we have two initial temperatures, $T_{i1}=5.0$ and
$T_{i2}=0.8$, and $T_f=0.4$.  If the evolution of the system is seen as
a slow evolution of the system towards parts of the configuration space
that correspond to lower free energies (deeper valleys), we can
reasonably expect that the system on its way from $T_{i1}$ to $T_f$
will visit valleys typical for $T_{i2}$. Hence, we expect that the
relaxation curves corresponding to $T_{i1}$ will, after some waiting
time $t_{(1,2)}$, essentially coincide with those obtained with a
starting temperature $T_{i2}$. A reasonable assumption is that the time
lag $t_{(1,2)}$ will be roughly of the order of magnitude of the
relaxation time $\tau_e(T_{i2})$ for an equilibrated system at
temperature $T_{i2}$.  In terms of the two time correlations, this
suggests a relation of the form $C^{(2)}_q(t_w, t+t_w) \simeq
C^{(1)}_q(t_w+t_{(1,2)}, t+t_w+t_{(1,2)})$. (Here the superscript
corresponds to the value of the starting temperature). This conjecture
is tested in Fig.~\ref{C2}, and is seen to be reasonably well borne out
by the numerical data in that the curve for $T_i=0.8$ ($\tau_e\approx
100$~\cite{Kob}) for $t_w=10$ is very similar to the curve for $T_i$=5.0,
$t_w=160$. The natural consequence of this behavior is that for a given
starting temperature $T_i$, we expect ``aging'' phenomena to take place
only for waiting times that exceed the equilibrium relaxation time
$\tau_e(T_i)$, whereas for $t_w<\tau_e(T_i)$ the relaxation behavior is
almost independent of $t_w$. In other words, it takes the system a time
of order $\tau_e(T_i)$ to realize that the quench has created a
nonequilibrium situation. In fact, we have observed that for an initial
temperature $T_i=0.466$,  for which $\tau_e \sim 10^5$, aging effects
such as those depicted by Fig.~\ref{C1}  are very weak on the time
scale of the simulation~\cite{barrat97}.

This situation could seem somewhat discouraging, in the sense that it
implies that the observation of interesting effects will require either
large values of the temperature jumps $T_i-T_f$, or simulations on time
scales much larger than $10^5$ time units. The question that immediately
arises if we consider large temperature jumps is to which extent the
aging  effects will display the universal behavior observed in real
experiments, where the typical parameters of the quench are very
different.  Experimentally, the most
striking observation, which is also predicted by several theoretical
models, is that the curves corresponding to different values of the
waiting time can be rescaled in the form 
\begin{equation}
C_q(t_w,t+t_w)= C_q^{st}(t) + C_q^{ag}\left({h(t+t_w)\over
h(t_w)}\right).  
\label{scaling_form} 
\end{equation} 
Here the first term corresponds to a short time dynamics that does not
depend on $t_w$, while the second term, or ``aging'' part, depends only
on the ratio $h(t_w+t)/h(t_w)$, were $h$ is an increasing function of
$t$. In many cases, $h(t)\simeq t$ (the so called ``simple aging''
case), or $h(t)\simeq t^\alpha$, so that the aging part is simply a
function of $t/t_w$. The existence of the $t_w$ independent, short time
part is evident from the data shown in Fig.~\ref{C1}. The scaling
assumption for long times is tested in Fig.~\ref{C1_rescaled}, for an
initial temperature $T_i=5.0$.  Except for the data that corresponds to
small values of $t_w$ ($t_w<10$), which -not surprisingly- do not fit
the scaling picture, the long time decay of the different curves can be
superimposed, indicating the validity of the scaling ansatz.  In
Fig.~\ref{ts_vs_tw}, the relaxation  time $t_r$ of $C_q(t_w,t+t_w)$
 is displayed as a function of the waiting time.  This relaxation time
was defined, somewhat arbitrarily, as the time it takes to reach the
value $C_q=0.45$, which is the lowest value of $C_q$ for $t_w=39810$.
For $T_i=5.0$ the two times turn out to be roughly proportional,
$t_r\sim t_w^{\alpha}$, $\alpha\approx 0.88$ (see Fig.~\ref{ts_vs_tw}),
which indicates that the system quenched from this $T_i$ approximately
follows the ``simple aging'' scaling over the time scales we are able
to investigate.

A similar analysis can be carried out for systems quenched from lower
temperatures, $T_i=0.8$ and $T_i=0.466$.  As we mentioned earlier,
aging effects in these systems start being appreciable only for  $t_w >
\tau_e(T_i)$ and thus $t_r$ is essentially constant for
$t_w<\tau_e(T_i)$ (see Fig.~\ref{ts_vs_tw}). For $t_w> \tau_e(T_i))$,
however, $t_w$ and $t_r$ are again roughly  proportional to each
other~\cite{barrat97}, as can be inferred from the corresponding curves
in Fig.~\ref{ts_vs_tw}.

A number of interesting conclusions and perspectives can be drawn from
these MD results on the nonequilibrium dynamics of a glass forming
liquid. The similarity with comparable studies on spin glasses, in
particular the existence of a ``universal'' $t/t_w$ scaling in the
aging behavior, is striking.  This can be seen as an indication that,
in spite of very different forms of the Hamiltonians and of the
microscopic dynamics, the geometry of phase space, which ultimately
determines the long time behavior, is not dissimilar in spin and
structural glasses.  Because of the short time scales that are
investigated in this work, we have carefully examined the role of
initial conditions. Their influence can be rationalized by introducing
an effective waiting time $t_w+\tau_e(T_i)$, for a system quenched
instantaneously from a temperature at which the equilibrium relaxation
time is $\tau_e(T_i)$. This notion might be useful in understanding
annealing experiments. Finally, the obvious extension of this work will
be to investigate the behavior of the one-particle response functions
under the same conditions and on the same time scales as was done in
this work. This would allow to investigate possible violations of the
fluctuation dissipation theorem during the aging process, and will be
the subject of future work~\cite{barrat97}.

\acknowledgments 
We benefited from useful discussions and correspondence with L.
Cugliandolo, J. Kurchan, A. Barrat, J.-P. Bouchaud and M. M\'ezard.
This work was supported by the Pole Scientifique de Mod\'elisation
Num\'erique at ENS-Lyon, and the Deutsche Forschungsgemeinschaft
through SFB 262.

\newpage
\section*{Figures}
\begin{figure}
\caption{Potential energy of the system as a function
of time for different values of $T_i$ (inset). Main figure:
$e_{pot}+7.17$ versus $t$, demonstrating that the time
dependence of $e_{pot}$ is compatible with a power-law.}
\label{energy_vs_t}
\end{figure}

\begin{figure}
\caption{$C_q(t_w,t+t_w)$ versus $t$ for $t_w=0$, 10, 100, 1000,
10000 and
39810 (from left to right); $T_i=5.0$, $T_f=0.4$, $q=7.2$.}
\label{C1}
\end{figure}

\begin{figure}
\caption{$C_q(t_w,t+t_w)$ for $T_f=0.4$ and  two different values of
$T_i$.
Thin lines: $T_i=0.8$, $\tau_e(0.8)\approx 100$ time units,
$t_w=0$, 10, 100, 1000, 10000 and 39810 (from left to right).
Dotted line: $T_i=5.0$, $t_w=160$.}
\label{C2}
\end{figure}

\begin{figure}
\caption{The data of Fig.~2, rescaled in such a way that all
curves coincide for $C_q=0.45$.}
\label{C1_rescaled}
\end{figure}

\begin{figure}
\caption{Relaxation time $t_r$ of $C_q(t_w,t+t_w)$ versus $t_w$ for
$T_f=0.4$ and three values of the initial temperature $T_i$. The
dashed line is a power-law with an exponent 0.882.}
\label{ts_vs_tw}
\end{figure}

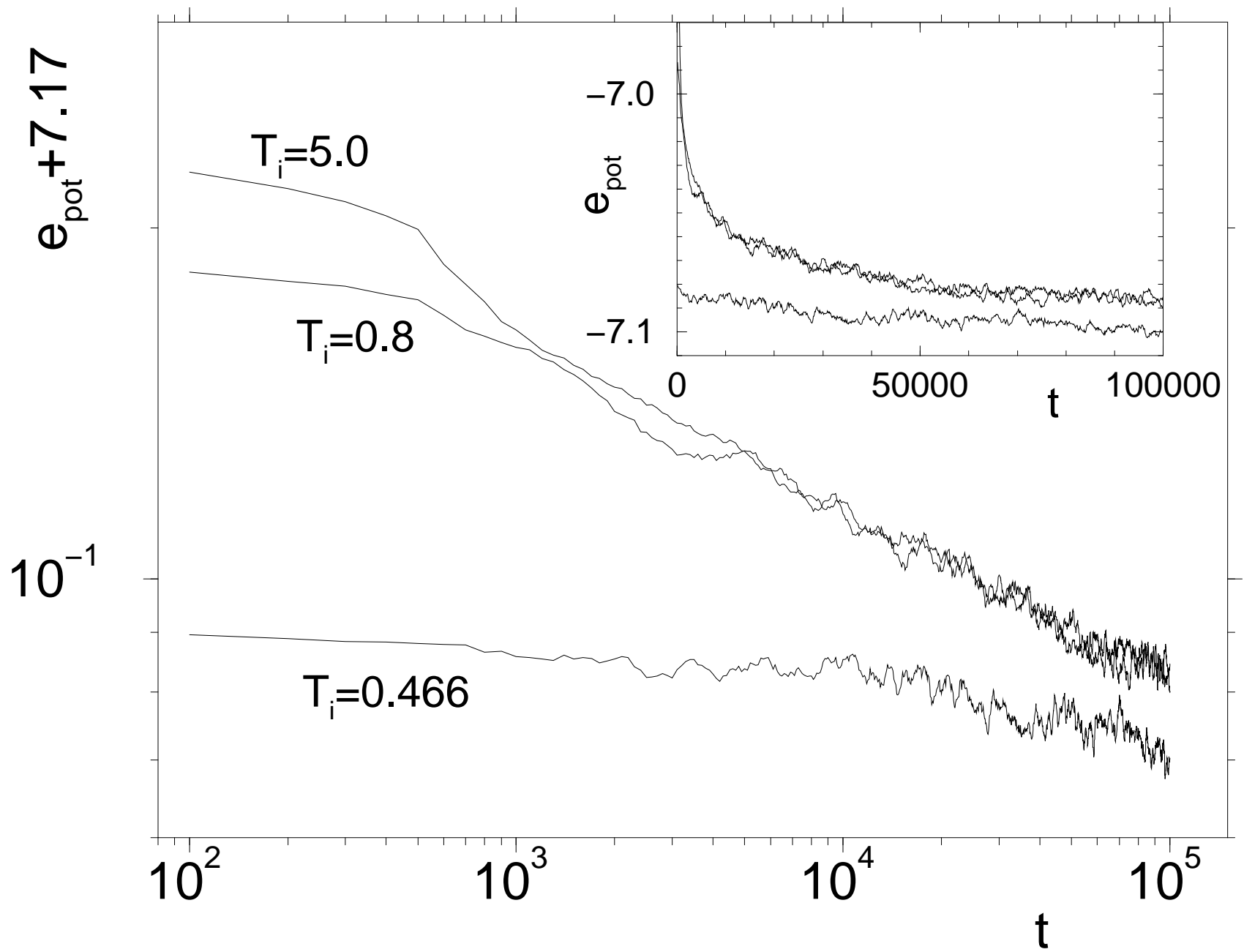
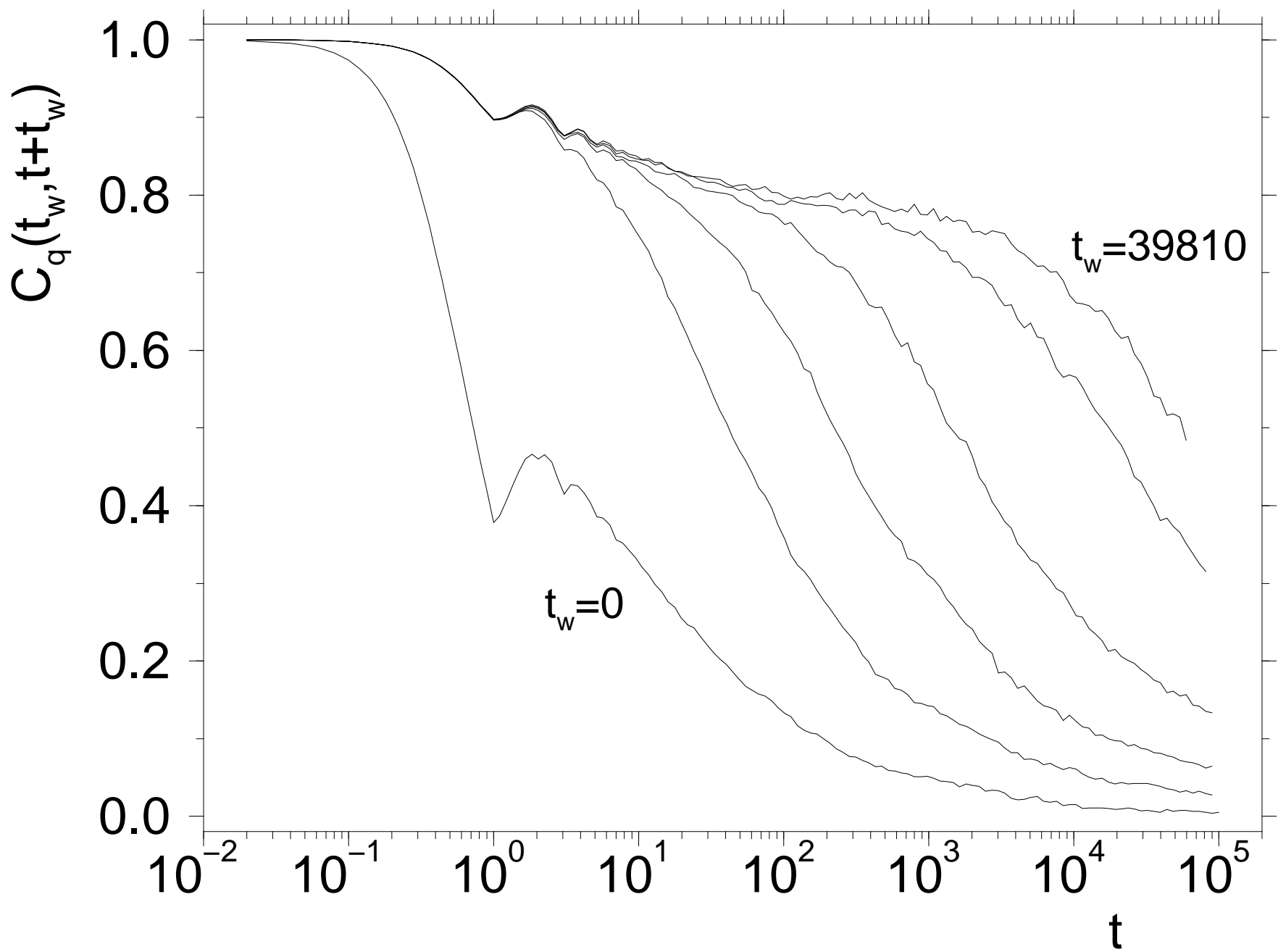
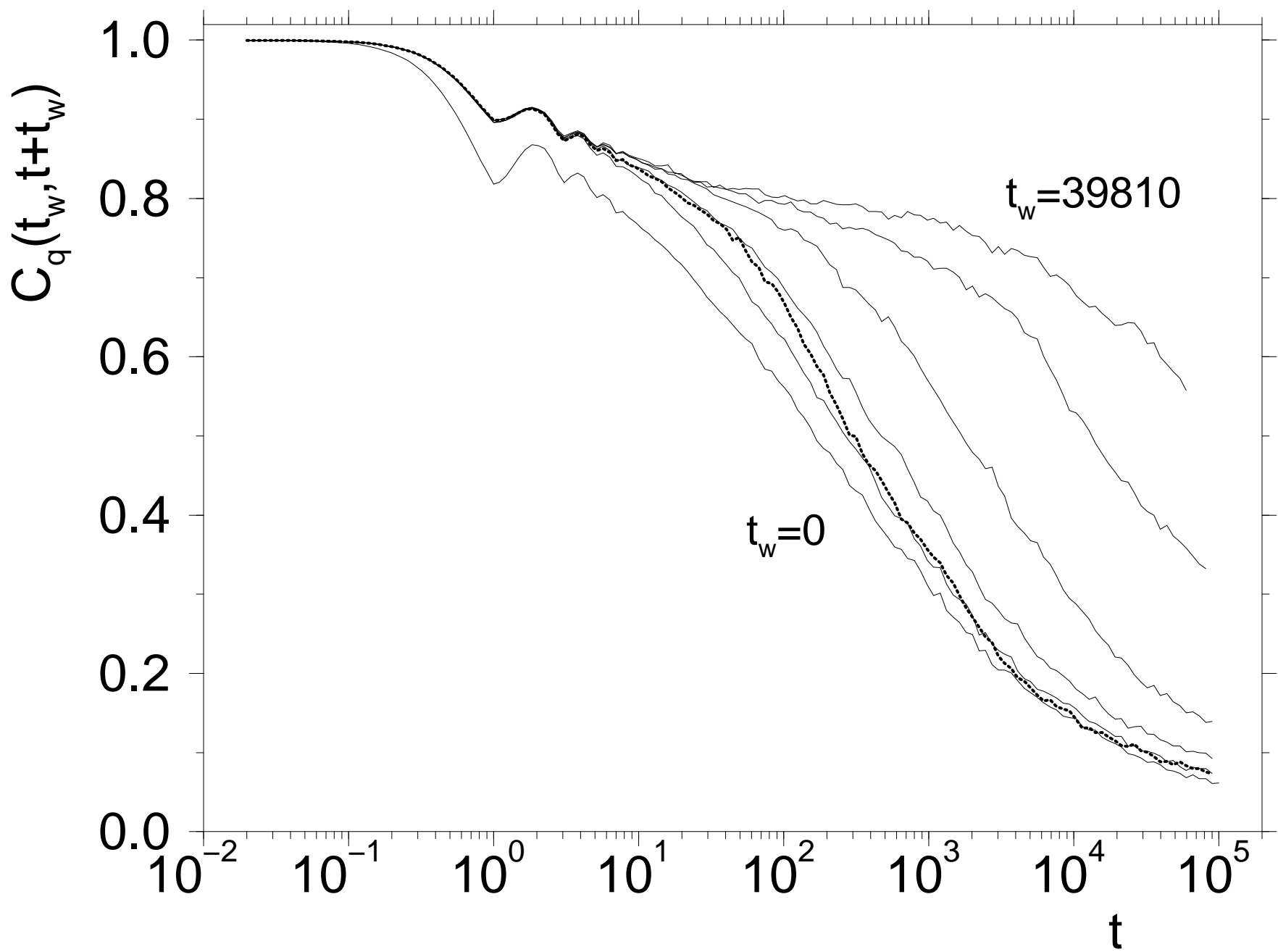
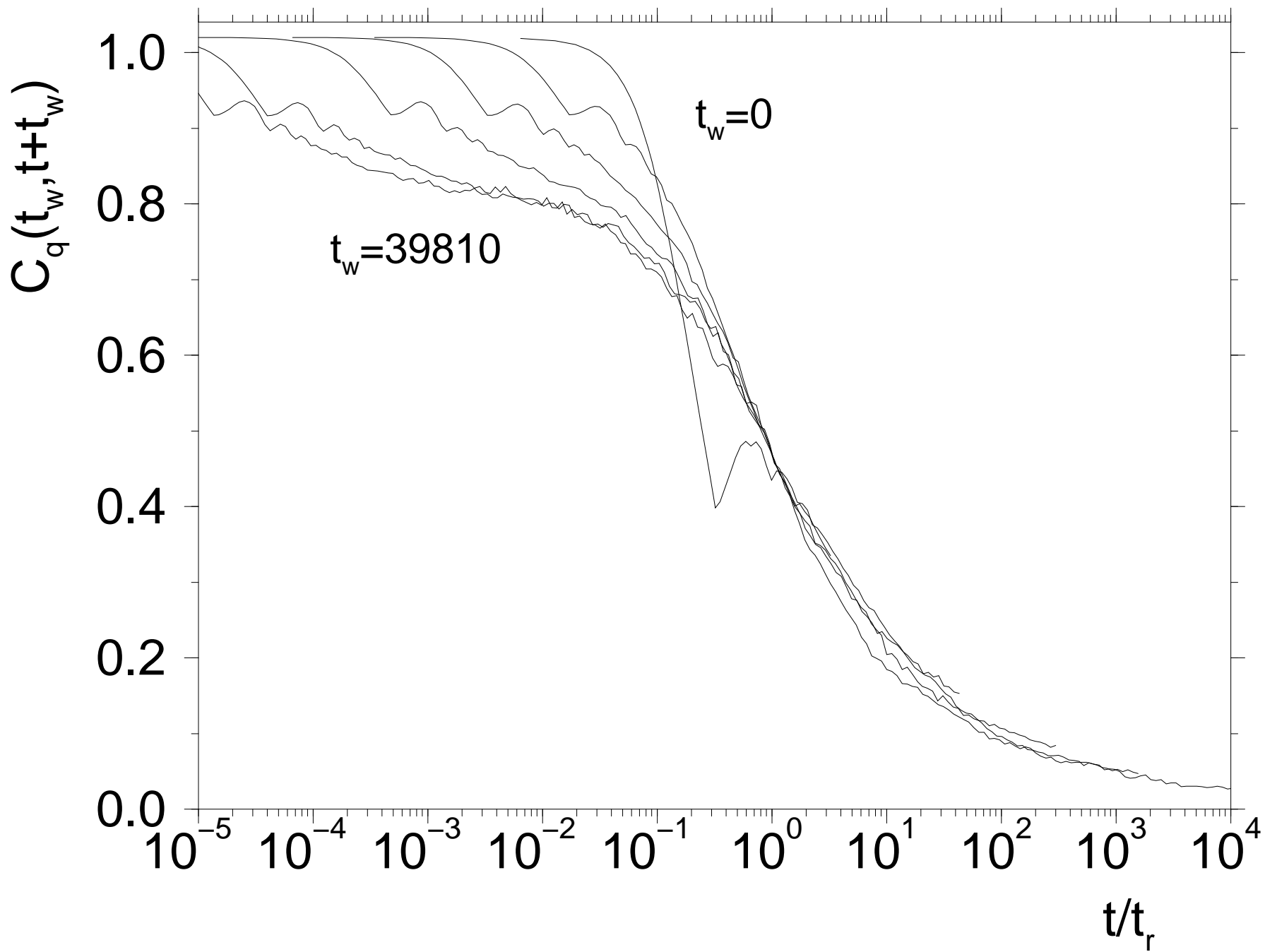
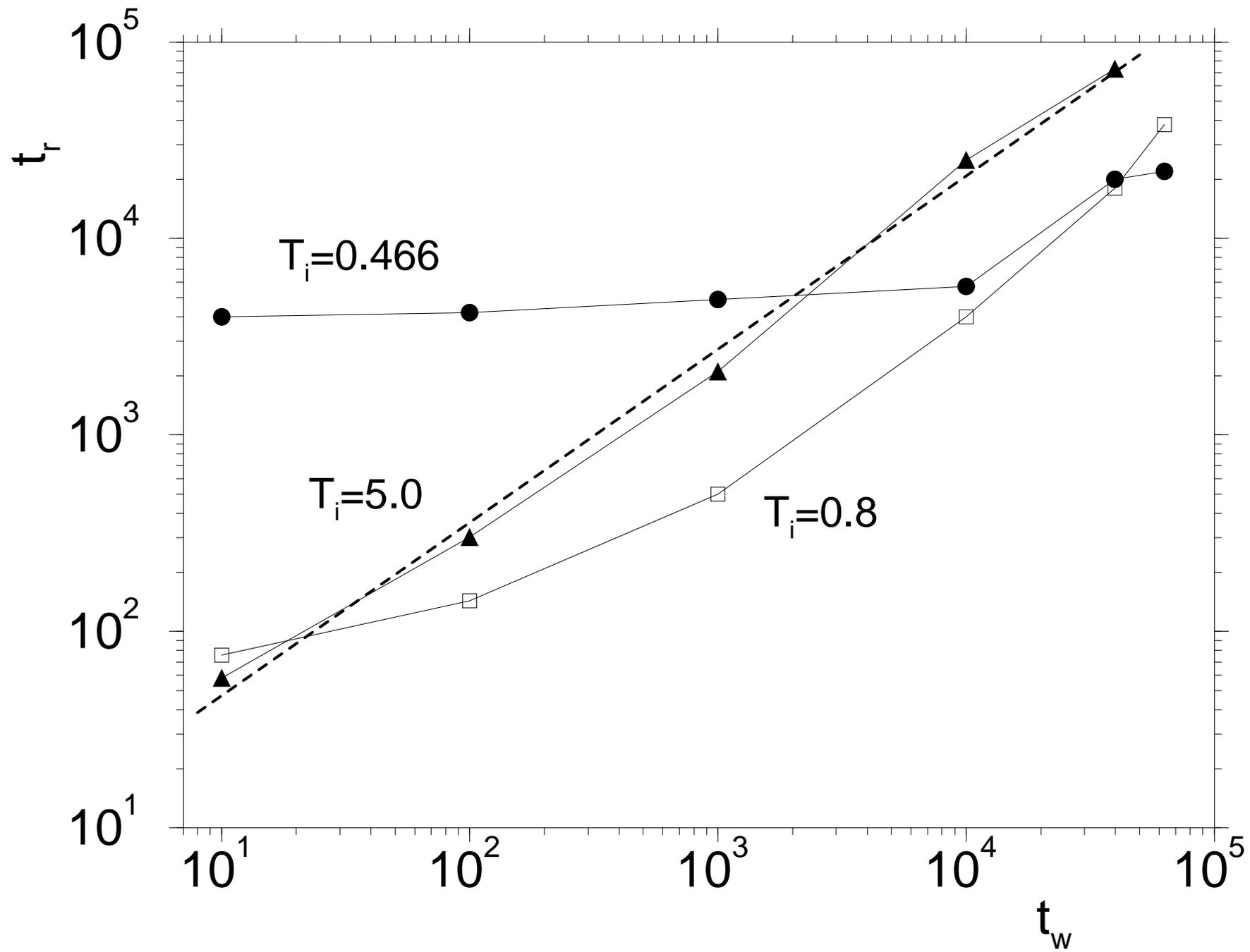
\end{document}